\documentclass[manuscript, screen, dvipsnames]{acmart}
\usepackage{multirow}
\usepackage{amsmath}
\usepackage[inline]{enumitem} 
\usepackage{subcaption}
\usepackage{xcolor}

 \usepackage{lipsum}
\usepackage{arydshln}
\makeatletter
\def\adl@drawiv#1#2#3{
        \hskip.5\tabcolsep
        \xleaders#3{#2.5\@tempdimb #1{1}#2.5\@tempdimb}%
                #2\z@ plus1fil minus1fil\relax
        \hskip.5\tabcolsep}
\newcommand{\cdashlinelr}[1]{%
  \noalign{\vskip\aboverulesep
          \global\let\@dashdrawstore\adl@draw
          \global\let\adl@draw\adl@drawiv}
  \cdashline{#1}
  \noalign{\global\let\adl@draw\@dashdrawstore
          \vskip\belowrulesep}}
\makeatother

\usepackage{tikz}
\usetikzlibrary{automata, arrows, bayesnet, bending}

\usepackage{tikzscale}

\copyrightyear{2023}
\acmYear{2023}
\setcopyright{acmlicensed}
\acmConference[RecSys '23]{Seventeenth ACM Conference on Recommender Systems}{September 18--22, 2023}{Singapore, Singapore}
\acmBooktitle{Seventeenth ACM Conference on Recommender Systems (RecSys '23), September 18--22, 2023, Singapore, Singapore}
\acmPrice{15.00}
\acmDOI{10.1145/3604915.3608777}
\acmISBN{979-8-4007-0241-9/23/09}

\pdfoutput=1

\begin{document}
\title{A Probabilistic Position Bias Model for Short-Video Recommendation Feeds}

\author{Olivier Jeunen}
\affiliation{
  \institution{ShareChat}
  \city{Edinburgh}
  \country{United Kingdom}
} 

\begin{abstract}
Modern web-based platforms often show ranked lists of recommendations to users, in an attempt to maximise user satisfaction or business metrics.
Typically, the goal of such systems boils down to maximising the exposure probability ---conversely, minimising the rank--- for items that are deemed ``\emph{reward-maximising}'' according to some metric of interest.
This general framing comprises music or movie streaming applications, as well as e-commerce, restaurant or job recommendations, and even web search.
\emph{Position bias} or \emph{user} models can be used to estimate exposure probabilities for each use-case, specifically tailored to \emph{how} users interact with the presented rankings.
A unifying factor in these diverse problem settings is that typically only one or several items will be engaged with (clicked, streamed, purchased, et cetera) before a user leaves the ranked list.

\emph{Short-video feeds} on social media platforms diverge from this general framing in several ways, most notably that users do not tend to leave the feed after, for example, \emph{liking} a post.
Indeed, seemingly infinite feeds invite users to scroll further down the ranked list.
For this reason, existing position bias or user models tend to fall short in such settings, as they do not accurately capture users' interaction modalities.
In this work, we propose a novel and probabilistically sound personalised position bias model for feed recommendations.
We focus on a 1\textsuperscript{st}-level feed in a hierarchical structure, where users may enter a 2\textsuperscript{nd}-level feed via any given 1\textsuperscript{st}-level item.
We posit that users come to the platform with a given \emph{scrolling budget} that is drawn according to a discrete power-law distribution, and show how the \emph{survival function} of said distribution can be used to obtain closed-form estimates for personalised exposure probabilities.
Empirical insights gained through data from a large-scale social media platform show how our probabilistic position bias model more accurately captures empirical exposure than existing models, and paves the way for improved \emph{unbiased} evaluation and learning-to-rank.
\end{abstract}

%



\maketitle

\section{Introduction \& Motivation}
Recommender system applications on the web often operate in a ranking fashion, showing ordered lists of items to users in an attempt to optimise some metric(s) of interest.
Such metrics typically reflect user satisfaction, business goals or fairness concerns.
With the ranking paradigm comes an important caveat: items that are shown at higher positions are more likely to be \emph{exposed} to the user, and this discrepancy should be taken into account when considering data from logged user interactions~\cite{Joachims2007}.
Indeed, it has implications for evaluation~\cite{Hofmann2014, Castells2022}, learning~\cite{Joachims2017}, and fairness of exposure~\cite{Diaz2020, Oosterhuis2021, Jeunen2021B}.
The problem of \emph{position bias} and its relevance to recommendation systems has been well-studied in recent years~\cite{Vardasbi2020, Chen2022, Ruffini2022, Oosterhuis2023}.
\looseness=-1

Most of these existing works focus on the classical Information Retrieval (IR) task of web search, where documents are ranked as search results to be surfaced for a given query.
Effective methods for de-biasing in web search are often transferable to recommendation domains, when we replace \emph{queries} with \emph{users} and \emph{documents} with \emph{items}.
As a result, evaluation metrics such as Normalised Discounted Cumulative Gain (nDCG) are a common choice when assessing top-$n$ recommendation quality~\cite{Valcarce2020}.
An often overlooked point is that the \emph{discount} is directly related to \emph{position bias}, and that well-chosen discount functions are necessary to consider nDCG an unbiased offline estimator of online reward~\cite{Jeunen2021thesis}.
This is a desirable feat, as discrepancies between off- and on-line evaluation results have plagued the recommender systems field for years~\cite{Beel2013, Garcin2014, Rossetti2016, Gilotte2018, Jeunen2018, Jeunen2019DS}.
Models of user behaviour, such as those underlying the rank-biased precision (RBP)~\cite{Moffat2008} or expected reciprocal rank (ERR)~\cite{Chapelle2009} metrics, can be used to construct discount functions for nDCG-like metrics that emulate the empirical position bias in a given system well.
Existing work in this area has largely focused on web search~\cite{chuklin2015click}, with extensions to general recommendation use-cases in e-commerce~\cite{Mei2022}.

\emph{Short-video feeds} on social media platforms, however, imply very different interaction paradigms than those prevalent in web search.
Indeed, users are unlikely to abandon the feed after, for example, \emph{liking} a post.
There is no \emph{information}, but rather an \emph{entertainment} need for users scrolling the feed.
As such, user models that are prevalent in other application areas are not directly applicable to our use-case~\cite{Zhang2020}.
Aside from more general work by \citeauthor{Wu2021}~\cite{Wu2021}, this topic has received relatively little research attention.
We specifically focus on a 1\textsuperscript{st}-level feed in a hierarchical structure, where users can either keep scrolling the current feed, or enter a ``\emph{more-like-this}'' 2\textsuperscript{nd}-level feed via any 1\textsuperscript{st}-level item.
Indeed, such user interfaces have gained popularity recently, and can be found on Reddit, Instagram and ShareChat, among others.
Our hypothesis is that users come to the platform with a ``\emph{scrolling budget}'', reflecting how far they are willing to scroll before abandoning the feed.
This \emph{budget} is personalised, context-dependent, and drawn from a discrete power-law distribution such as the Yule-Simon distribution with shape parameter $\rho$~\cite{Yule1925, Simon1955}.
Figure~\ref{fig:yulesimon}(a) visualises how this family of distributions can represent a wide variety of stochastic budgets and, hence, scrolling behaviours.

We show how the survival function of this distribution can be used to obtain closed-form estimates for personalised exposure probabilities that have a sound theoretical basis, and show how they pave the way for improved unbiased evaluation and learning-to-rank in feed recommendation settings.

The main contributions we present in this paper are the following:
\begin{enumerate}
\item We propose a novel Contextual, Personalised, Probabilistic POsition bias model for feed recommendations: \texttt{C-3PO}.
\item We empirically validate using real-world data that \texttt{C-3PO} is better able to capture exposure probabilities than existing methods, whilst having a stronger theoretical basis.
\item We show how \texttt{C-3PO} can be used for improved unbiased evaluation and learning in feed ranking scenarios.
\end{enumerate}
\section{Methodology \& Contributions}\label{sec:contribution}
\subsection{Position Bias Models}\label{sec:background}
For ease of terminology but without loss of generality, assume we want to maximise the \emph{quality} ($Q$) of items that are viewed ($V$) by a user in a certain \emph{context} ($X$).
\footnote{Note that in classical web-search use-cases, \emph{view} events $V$ are unobserved, as multiple items are presented to the user simultaneously~\cite{Joachims2005}.
This is different from our use-case, where items impressed to the user take up most of their mobile screen, and we can extract labels $V$ from scrolling behaviour.}
As is typical, the true \emph{quality} is not an observable quantity, but we can estimate it from logged user interactions.
We will refer to these as \emph{clicks} ($C$) that occur when a post is both viewed and deemed quality, but they can represent more general engagement (e.g. \emph{likes}).
A ranking policy $\pi$ is in place, deciding at which \emph{rank} ($R$) a given post will appear, dependent on  the context $X$ and a quality estimate $\widehat{Q}$.
We can describe the relation between the unobservable quality $Q$ and the observable quantities $C$, $V$ and $R$ as follows (omitting item $i$):
\begin{equation}
    \mathsf{P}(Q|X) = \frac{\mathsf{P}(C|R,X)}{\mathsf{P}(V|R,X,\pi)}.
\end{equation}
This is a well-known result, showing how we can obtain unbiased estimates of quality conditional on context, by reweighting observed clicks with exposure or viewing probabilities (i.e. \emph{inverse propensity scoring}, or IPS~\cite[Ch. 9]{Owen2013}).
In this work, we do not focus on \emph{selection bias}, i.e. bias stemming from the ranking policy $\pi$ influencing the rankings~\cite{Ovasisi2020,Oosterhuis2020}.
Instead, we wish to estimate the \emph{causal effect} of rank on exposure, conditional on context.

\begin{figure}
\includegraphics{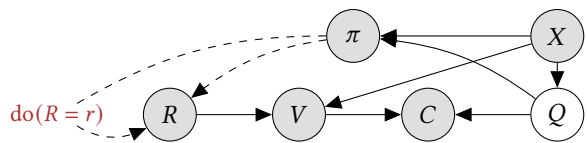}
    \caption{Probabilistic Graphical Model (PGM) detailing how interventional data collection allows for unbiased estimation of the causal effect between the rank $R$ and view events $V$, by removing incoming causal edges from the deployed ranking policy $\pi$.}
    \label{fig:PGM}
\end{figure}

Without considering contextual information, \citeauthor{Joachims2017} originally proposed to perform randomised interventions on the rank of a given item to obtain such estimates~\cite{Joachims2017}.
Further extensions proposed to leverage historical interventions stemming from natural experiments~\cite{Agarwal2019}, and this was further extended to be context-dependent~\cite{Fang2019}.
Whichever of these methods is adopted, we essentially obtain data from the interventional distribution describing $\mathsf{P}(V|{\rm do}(R=r), X)$, with the ${\rm do}$-operator following \citeauthor{Pearl2009}'s seminal work~\cite{Pearl2009}.
These interventions remove the dependency of empirical views on the deployed ranking policy $\pi$, and thus:
\begin{equation}
    \mathsf{P}(Q|X) = \frac{\mathsf{P}(C|{\rm do}(R),X)}{\mathsf{P}(V|{\rm do}(R),X)}.
\end{equation}
We visualise this interventional procedure with the Probabilistic Graphical Model (PGM) shown in Figure~\ref{fig:PGM} --- this causal view is often left implicit in related work.
The main benefit from this derivation is that it allows us to obtain an unbiased estimate of item quality from observable quantities alone, but downstream applications of accurate exposure probabilities are threefold:
\begin{enumerate*}
    \item ranking policies can be evaluated more reliably with \emph{offline} estimators of \emph{online} metrics that depend on exposure~\cite{Joachims2017},
    \item ranking policies can be learnt to maximise better objectives by replacing the observed labels $C$ with estimates of quality (essentially \emph{de-biasing} them)~\cite{Oosterhuis2020}, and
    \item fairness metrics related to ``\emph{equity of exposure}'' among content creators, for example, can be estimated and optimised more reliably~\cite{Diaz2020,Jeunen2021B}.
\end{enumerate*}

Note that (1) is essentially what the DCG metric aims to do, by discounting the ``\emph{gain}'' at every rank (independent of context).
The standard inverse-logarithmic discount function for DCG is commonly adopted ($\mathsf{P}_{\rm dcg}$ in Eq.~\ref{eq:dcg}), and forms the basis for ranking evaluation in the wider IR and recommendation field.
Nevertheless, if the exposure probabilities implied by this discount function are inaccurate, DCG fails to be an unbiased estimator of online gains.
If this is the case, we can consider other parametric discount functions, with several forms.
In line with DCG, it is natural to consider an inverse logarithmically shaped function (indicating diminishing position bias effects at lower ranks), or an exponential decay function (in line with the user model adopted by RBP~\cite{Moffat2008}).
These families of functions will be the baselines for our experiments, parameterised by $\alpha \in \mathbb{R}^{+}_{0}$ and $\gamma \in [0,1]$ respectively: \footnote{We interpret them as probabilities, but these discounts were not originally motivated by sound probabilistic models of user behaviour (bar RBP~\cite{Moffat2008}).}
\begin{equation}\label{eq:dcg}
\mathsf{P}_{\rm dcg}(V=1|R=r) = \frac{1}{\log_{2}(r+1)},\qquad\quad
\mathsf{P}_{\log}(V=1|R=r) = \frac{1}{\ln(e+\alpha (r-1))}, \qquad\quad
\mathsf{P}_{\exp}(V=1|R=r) = \gamma^{r-1}.
\end{equation}
In the context of web search applications, regression- and deep neural network (DNN)-based position bias models have been proposed as well~\cite{Wang2018PB,Ai2018}.
Whilst effective, there are practical drawbacks to implementing such methods in real-world applications:
\begin{enumerate*}
    \item it is computationally intensive to obtain $\mathsf{P}(V|R)$ from a forward pass in a neural network,
    \item it requires significant engineering effort to make such models accessible when performing e.g. offline evaluation or exposure fairness analyses,
    \item DNNs do not guarantee robust estimates in low-data regimes and fail to encode desirable properties such as monotonicity of $\mathsf{P}(V|R)$ w.r.t. $R$, and
    \item the black-box nature of these models does not allow us to obtain an improved understanding of \emph{how} users interact with the rankings they are presented with.
\end{enumerate*}
For these reasons, we place simple models at the focal point of our work, with a minimal amount of learnable parameters.

Finally, note that all of the aforementioned models are \emph{independent} of contextual information $X$.
\citeauthor{Fang2019} propose a DNN-based model whose output reflects $\mathsf{P}(V|R,X)$ instead; adopting a classical Multi-Layer Perceptron (MLP) architecture while conjecturing that architectural improvements could improve results~\cite{Fang2019}.
Naturally, their method suffers from the same drawbacks as non-contextual deep models.
\citeauthor{Wu2021} leverage Gradient-Boosted Decision Trees for an e-commerce feed recommendation application, with similar concerns for practical implementations~\cite{Wu2021}.
We can address some of these shortcomings by viewing position bias through a probabilistic lens, as it provides a theoretically sound basis to reason about exposure, allowing for more sample- and parameter-efficient learning.

Next to the Position-Based Model, alternative models of user behaviour have been proposed in the web search literature~\cite{Zhang2020}.
Such methods require us to model the probability that a user continues down a ranked list, conditional on the items viewed so far.
As this significantly increases modelling complexity, it is out of scope for this short article.

\subsection{Probabilistic Position Bias Models}
The intuitive logarithmic discount originally proposed by \citeauthor{Jarvelin2002} has been widely adopted in the research literature~\cite{Jarvelin2002}, along with the exponential form that motivates RBP~\cite{Moffat2008}, or cascade-based alternatives~\cite{Craswell2008,Chapelle2009DBN}.
Model-based approaches~\cite{Wang2018PB,Ai2018,Wu2021,Fang2019} hold promise to move beyond mere intuition, but their implementation has several practical drawbacks that hinder widespread adoption. 
We adopt the Contextual Position-Based Model (CPBM) proposed by \citeauthor{Fang2019}~\cite{Fang2019}, but aim to tackle the position bias estimation problem through a probabilistic lens.

We introduce an additional random variable $D$, referring to ``\emph{scroll depth}''.
In our hierarchical use-case, users scroll through ranked posts on the 1\textsuperscript{st}-level feed until they decide to either enter a 2\textsuperscript{nd}-level feed via an item of their liking, or decide to abandon the feed altogether.
A large majority of sessions sees users entering a 2\textsuperscript{nd}-level feed via a highly ranked post, fewer sessions lead to 2\textsuperscript{nd}-level feeds at lower ranks, and a small minority sees users abandoning the feed after scrolling further.
As such, our discrete \emph{scroll depth} random variable follows a power-law distribution, such as the Yule-Simon distribution visualised in Figure~\ref{fig:yulesimon}(a)~\cite{Yule1925,Simon1955}.
This distribution includes a shape parameter $\rho \in \mathbb{R}^{+}_{0}$.
When ${\rm B}(\cdot,\cdot)$ represents the Beta function, its Probability Mass Function (PMF) at depth $D=d$ is given by:
\begin{equation}
    \mathsf{P}_{{\rm Yule-Simon}(\rho)}(D=d) = \rho {\rm B} (d, \rho+1).
\end{equation}
Having defined a distribution for scroll depth, we can define the relationship between scroll depth and position bias.
Indeed, a post ranked at position $R=r$ will be viewed if and only if the user decides to scroll at least up until that rank.
This implies a negative relationship with the Cumulative Distribution Function (CDF) of $D$:
\begin{equation}
    \mathsf{P}(V=1|R=r) = \mathsf{P}(D \geq r) = 1 - \mathsf{P}(D < r).
\end{equation}
That is, the position bias at rank $r$ can be derived from the survival function of the scroll depth distribution.
For the Yule-Simon($\rho$)-distribution, this quantity is given by:
\begin{equation}
\mathsf{P}_{{\rm prob}}(V=1|R=r) =
\begin{cases}
      1 & \text{if}~r=1,\\
      (r-1) {\rm B} (r-1, \rho+1) & \text{otherwise.}
\end{cases} \label{eq:prob}
\end{equation}
This derivation gives rise to a probabilistic position bias model that is theoretically sound, and motivated by a plausible model of user behaviour, specifically tailored to the use-case of feed recommendations.
Note that we adopt the Yule-Simon distribution largely for illustratory purposes in this Section, but that our analysis remains general.
As such, we could adopt more involved probability distributions for scroll depth to reflect position bias curves that would be unrealisable by Yule-Simon.
A promising alternative here is the generalised Waring distribution, as it can reflect a wider set of position bias curves~\cite[\S 6.2.3]{johnson2005univariate}.
As this probability distribution is defined by multiple parameters, it is out-of-scope for the purposes of this short article.
Even though the probabilistic position bias model proposed so far is neither contextual nor personalised, we expect the position bias curves emanating from Eq.~\ref{eq:prob} to be more aligned with empirical biases in feed recommendation scenarios than those produced by the classical approximations in Eq.~\ref{eq:dcg}.

We visualise the position bias curves that emanate from our proposed model in Figure~\ref{fig:yulesimon}(b), contrasting them with those arising from classical approximations.
Indeed, whilst we observe that $\mathsf{P}_{\rm dcg}$ does not discount aggressively enough, both $\mathsf{P}_{\rm log}$ and $\mathsf{P}_{\rm exp}$ cannot accurately represent similar position bias curves as our probabilistically inspired method.

\subsubsection{Connections to Mean Reciprocal Rank (MRR)~\cite{Voorhees1999}}
\begin{figure}
    \centering
    \includegraphics[width=\linewidth]{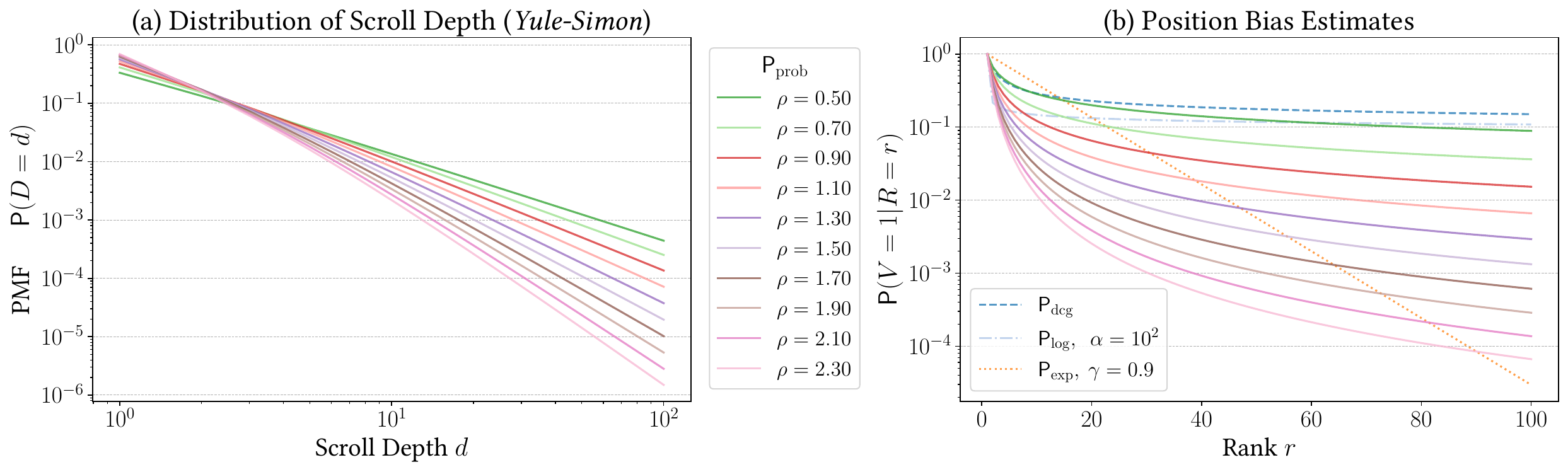}
    \caption{Visualising the distributions induced by our proposed probabilistically inspired position bias models.\\
    \textbf{Left (a)}: the Probability Mass Function (PMF) for the discrete Yule-Simon distribution with varying shape parameter $\rho$, at different scroll depth levels. By personalising the parameter $\rho$, we can express a wide breadth of distributions that represent scrolling behaviour.\\
    \textbf{Right (b)}: the position bias curves that emanate from our probabilistically inspired position bias model, with varying shape parameter $\rho$, along with common traditional methods.
    Our proposed model captures scrolling behaviour that is different to existing methods.
    }
    \label{fig:yulesimon}
\end{figure}

A commonly used offline evaluation metric in the web search domain is MRR, which averages the reciprocal rank of the first relevant document in every ranking.
When assuming binary relevance labels and a single relevant item per ranking, this metric is equivalent to DCG where the discount corresponds to the reciprocal rank: $\mathsf{P}_{\rm rr}(V=1|R=r) = \frac{1}{r}$.
Recent work in the e-commerce recommendation domain highlights how this discount function aligns much more closely with the empirical position bias on their platform, and reports improved alignment between offline and online evaluation metrics~\cite{Mei2022}.
There is an intricate connection between the reciprocal rank discount and our proposed method, when we consider the Yule-Simon distribution with parameter $\rho=1$.
To see this, recall that the Beta function can be written in terms of Gamma functions as ${\rm B}(x,y) = \frac{\Gamma(x)\Gamma(y)}{\Gamma(x+y)}$, and $\Gamma(x)=(x-1)!$ for positive integers $x$.
Then, consider for $r>1$:
\begin{equation}
\mathsf{P}_{\rm prob}(V=1|R=r) = (r-1) {\rm B}(r-1, 2) =  (r-1) \frac{\Gamma(r-1)\Gamma(2)}{\Gamma(r-1+2)} = (r-1) \frac{\Gamma(r-1)}{\Gamma(r+1)} = \frac{(r-1)}{(r-1)r} = \frac{1}{r} = \mathsf{P}_{\rm rr}(V=1|R=r). \qed
\end{equation}
As such, the specific discount function that gives rise to MRR can be seen as a special case of our proposed method, when we adopt the Yule-Simon distribution with parameter $\rho=1$.
This is reassuring, given MRR's strong track record in IR research and its recent successes~\cite{Mei2022}.
Our probabilistic model for scroll depth gives rise to a theoretically sound motivation for MRR.
Furthermore, the flexibility of our framework allows us to adopt alternative probability distributions with varying parameterisations, giving rise to a rich family of position bias curves, as shown in Figure~\ref{fig:yulesimon}(b).

Existing work has connected RR-based metrics to \emph{cascade} user models in web search use-cases with graded relevance labels~\cite{Chapelle2009}.
Their derivation still holds when replacing $\mathsf{P}_{\rm rr}$ with the more general and parameterisable $\mathsf{P}_{\rm prob}$~\cite[Def. 1]{Chapelle2009}.

\subsubsection{Contextual, Personalised, Probabilistic Position Bias Models: \texttt{C-3PO}}\label{sec:learning}
We wish to incorporate contextual information encoded in $X$, to obtain a \emph{conditional} probability distribution for the scroll depth: $\mathsf{P}(D|X)$.
To this end, we aim to learn a function that maps $X$ to the parameters that define $D$'s distribution.
In the case of Yule-Simon, this means we wish to learn a parameterised function $\rho_{\theta}(X)$ s.t. the implied conditional probability distribution for the position bias model maximises the likelihood of observed data $\mathcal{D}$:
\begin{equation}
        \prod_{(v,r,x) \in \mathcal{D}} \mathsf{P}(V=v|R=r,X=x).
\end{equation}
As is common, we minimise negative log-likelihood (NLL) instead:
\begin{equation}\label{eq:NLL}
        \ell\left(\widehat{\mathsf{P}}; \mathcal{D}\right) =  - \frac{1}{|\mathcal{D}|}\sum_{(v,r,x) \in \mathcal{D}} \log\left(\widehat{\mathsf{P}}(V=v|R=r,X=x)\right),
\end{equation}
where $\widehat{\mathsf{P}}$ denotes the estimated position bias model.
For the Yule-Simon distribution, we plug in the position bias formula from Eq.~\ref{eq:prob} with the learnt function $\rho_{\theta}(x)$ to obtain:
\begin{equation}
    \widehat{\mathsf{P}}_{\rm prob}(V=1|R=r,X=x) = 
\begin{cases}
      1 & \text{if}~r=1,\\
      (r-1) {\rm B} (r-1, \rho_{\theta}(x)+1) & \text{otherwise.}
\end{cases}
\end{equation}
Using standard gradient descent techniques, we can now aim to learn the optimal parameters $\theta$ for $\rho_{\theta}(x)$ that minimise the NLL in Eq.~\ref{eq:NLL}.
Note that we can introduce rank-based weights or cut-offs to, for example, up-weight higher positions in the list.
When the position bias model has limited capacity, this approach can help to optimise the accuracy of the model in typical top-$K$ scenarios.
Indeed, the lower exposure probabilities at the bottom of the rankings are easier to model, and this can drown out the importance of the higher positions.
We denote such cut-offs with NLL@$K$.

Finally, note that the optimisation procedure derived in this Subsection does not solely restrict itself to our probabilistically motivated position bias model.
Indeed, we can learn the parameters for the logarithmic and exponential forms in Eq.~\ref{eq:dcg} that minimise the NLL of the resulting distributions just as well, which we do in our experiments.

\section{Experimental Results \& Discussion}\label{sec:experiments}
We wish to answer the following research questions experimentally:
\begin{description}
\item[\textbf{RQ1}] \textit{Is our proposed probabilistic method able to model exposure probabilities more accurately than existing methods?}
\item[\textbf{RQ2}] \textit{Can the model leverage contextual signals effectively?}
\item[\textbf{RQ3}] \textit{Are the obtained position biases useful for downstream tasks, such as unbiased offline evaluation?}
\end{description}

Naturally, position biases are heavily influenced by specific use-cases, platforms and interface choices.
The methods we propose in this work are motivated by a short-video feed recommendation use-case, and even though our proposed framework is generally applicable, we expect the Yule-Simon instantiation to only hold merit in similar use-cases.

In order to empirically validate the performance of both our and earlier proposed methods, we require \emph{interventional} data with logged views $V$, ranks $R$, and contexts $X$.
To the best of our knowledge and at the time of writing, we are unaware of any such datasets being publicly available.
Existing Learning-to-Rank (LTR) datasets do not contain rank interventions and deal with web search use-cases, which imply very different modalities to ours.
For this reason, we need to resort to proprietary datasets, but additionally release an open-source Jupyter notebook to reproduce the position bias curves visualised in Figure~\ref{fig:yulesimon} at \href{https://github.com/olivierjeunen/C-3PO-recsys-2023}{github.com/olivierjeunen/C-3PO-recsys-2023}.

\subsection{Estimating Exposure Probabilities (RQ1--2)}
We obtain a sample of 1 million sessions of feed view events on a social media platform, where rank interventions occurred following Fig.~\ref{fig:PGM}, collected over five days in February 2023.
We perform an 80-20\% train-test split, aiming to predict whether recommendations were viewed based on their rank and contextual information.

We compare several non-contextual variants: the standard DCG discount function as well as the logarithmic and exponential forms in Eq.~\ref{eq:dcg}, and the probabilistic method based on the Yule-Simon distribution introduced in Eq.~\ref{eq:prob}.
The latter three methods include a single parameter ($\alpha, \gamma, \rho$ respectively), which we learn to minimise NLL@$K$ on the training set, following the procedure laid out in \S\ref{sec:learning}.
We implement this in Python 3.9 with the SciPy library~\cite{Virtanen2020}.

As an additional baseline, we include a non-parametric method that predicts the empirical average from the training data.
This approach should be expected to outperform the aforementioned methods, but it requires a hard-coded probability at every rank instead of the single parameter that the logarithmic, exponential, or probabilistic forms require.
Additionally, this approach cannot easily be extended to incorporate contextual information $X$.

For the contextual case, we adopt a single continuous user-based feature describing users' past average scroll depth, as well as a single continuous context-based feature, describing average scroll depth at the time of day. 
We adopt a simple linear model to estimate the distribution parameter from this input $X$: $\rho_{\theta}(x) = \theta^{\intercal}x$.
The functional forms for the parameters $\alpha$ and $\gamma$ are analogous.
As such, the contextual and personalised methods consist of only \emph{three} parameters each (assuming $x$ includes a constant 1-feature, emulating a bias term in $\theta$).
Even in this simplistic scenario, the contextual and personalised methods significantly outperform those that do not consider this information, as shown in Table~\ref{tab:results}.
Our contextual, personalised, probabilistic position bias model \texttt{C-3PO} achieves the lowest NLL@$K$ for a wide range of $K$, whilst requiring a minimum of learnable parameters or computing resources.
This yields a desirable trade-off between parsimony and model expressiveness when compared to complex model classes like neural networks (which would typically require orders of magnitude more parameters). 
We observe that this additionally allows us to be sample-efficient, as our method already performs well with only $\mathcal{O}(10^{4})$ samples.
Indeed, instead of modelling the entire curve at every possible value of $r$, our proposed method outputs a single scalar which can be used to obtain position bias estimates for all natural numbers.
The inductive bias we enjoy from well-motivated mathematical models greatly improves the methods' real-world usability, when compared to neural-network based alternatives.

\begin{table}[t]
    \centering
    \begin{tabular}{lcccccc}
    \toprule
    \multirow{2}{*}{\textbf{Model}} & \multicolumn{5}{c}{\textbf{Negative Log-Likelihood (NLL)}}\\
     ~& \textbf{@5} & \textbf{@10} & \textbf{@25} & \textbf{@50} & \textbf{@100} \\
    \cline{2-6}

    $\widehat{\mathsf{P}}_{{\rm dcg}}(V|R)$ & 0.5453 & 0.6320 & 0.5998 & 0.4973 & 0.3763\\
    $\widehat{\mathsf{P}}_{\log}(V|R)$ & \underline{0.5159} & \underline{0.6001} & 0.5900 & 0.5036 & 0.3833\\
    $\widehat{\mathsf{P}}_{\exp}(V|R)$ & 0.5202 & 0.6158 & 0.6089 & 0.5101 & 0.3673\\
    $\widehat{\mathsf{P}}_{\rm prob}(V|R)$ & \underline{0.5162} & \underline{0.6002} &\underline{0.5873} & \underline{0.4891} & \underline{0.3495}\vspace{1ex}\\
    \cdashline{1-6}
    \vspace{-2ex}~\\
    $\widehat{\mathsf{P}}_{{\rm empirical}}(V|R)$ & 0.5157 & 0.5999 & 0.5843 & 0.4813 & 0.3369\vspace{1ex}\\
    \cdashline{1-6}
    \vspace{-2ex}~\\
    $\widehat{\mathsf{P}}_{\log}(V|R,X)$ & \textbf{0.4852} & \textbf{0.5620} & 0.5577 & 0.4806 & 0.3555\\ 
    $\widehat{\mathsf{P}}_{\exp}(V|R,X)$ & 0.4883 & 0.5761 & 0.5778 & 0.4959 & 0.3652\\ 
    $\widehat{\mathsf{P}}_{\rm prob}(V|R,X)$ & \textbf{0.4850} & \textbf{0.5620} &\textbf{0.5551} & \textbf{0.4651} & \textbf{0.3325}\\ 
\bottomrule
    \end{tabular}
    \caption{NLL for position bias models on observed data, lower is better.
    The top-group are independent of contextual information, the middle baseline is a non-parametric method that predicts a sample average, the bottom-group include three parameters that were optimised via linear regression. 
    Marked fields indicate stat. sig. improvements over other methods in the same group at a 99\% level.}
    \label{tab:results}
\end{table}

\subsection{Unbiased Offline Evaluation (RQ3)}
The main task position bias models need to perform, is to deliver offline estimates of online performance.
Given a dataset of logged impressions $\mathcal{D}\coloneqq \{(x_{i}, a_{i}, r_{i}, c_{i})_{i=1}^{N}\}$ (contexts, actions, ranks, rewards) , we wish to estimate the expected reward we would have obtained under some different ranking policy $\pi$.
This policy maps a context $X$ and set of \emph{candidate} items $\mathcal{A}_{x}$ to a ranked list.
We will denote with the shorthand notation $\pi(a|x)$ the rank that item $a$ will be placed at when $\pi$ is presented with context $x$ (assuming $\mathcal{A}_{x}$ given).
Note that this framing is easily extended to more general stochastic ranking policies~\cite{Oosterhuis2021}.
Then, a dataset $\mathcal{D}$ and position bias model $\widehat{\mathsf{P}}$ can be used to to obtain an unbiased estimate of the reward we would obtain under $\pi$:
\vspace{-1ex}
\begin{equation}\label{eq:unbiased_dcg}
    \mathop{\mathbb{E}}\limits_{r \sim \pi}[C] \stackrel{1}{\approx} {\rm DCG}_{\widehat{\mathsf{P}}}(\mathcal{D}, \pi) \stackrel{2}{\approx}
    \frac{1}{N}\sum_{i=1}^{N}
    c_{i} \cdot \frac{\widehat{\mathsf{P}}(V=1|R=\pi(a_{i}|x_{i}), X=x_{i})}{\widehat{\mathsf{P}}(V=1|R=r_{i}, X=x_{i})} .
\end{equation}
Here, the first approximation $\stackrel{1}{\approx}$ is due to the inherent assumptions of the DCG metric (compared to, e.g., cascade-based alternatives), whereas the second only exists because we resort to an empirical average over the observed data $\mathcal{D}$ and estimated position biases via $\widehat{\mathsf{P}}$.
Assuming unbiasedness of $\widehat{\mathsf{P}}$, the unbiasedness of the metric in Eq.~\ref{eq:unbiased_dcg} is easily recognised, as it is an application of importance sampling or IPS~\cite{Owen2013}.
As is typical for IPS-based methods, techniques like capping or introducing control variates can improve their finite-sample performance by reducing variance~\cite{Gilotte2018, Swaminathan2015snips}.
We do not consider such extensions in this short article, but remark that they are likely to further improve performance.

To validate the utility of these offline estimates, we perform an online experiment on a social media platform that operates a short-video recommendation feed.
Thus, we obtain samples from the reward distribution by sampling $\mathbb{E}_{r \sim \pi}[C]$ directly and taking an empirical average per day, for five days.
Then, for varying context-independent position bias models (optimised $@100$), we obtain offline estimates of online reward via Eq.~\ref{eq:unbiased_dcg}, and evaluate the offline estimates by Pearson's correlation coefficient between the ground truth and the offline estimate, over 5 days.

Table~\ref{tab:results2} shows relative improvements in correlation over the classical DCG formulation.
We observe that our probabilistically motivated position bias model is able to significantly improve the offline-online correlation compared to existing methods, and conjecture that the context-dependent variant can lead to further improvements.
This highlights the importance of a well-motivated position bias model, and is a strong argument in favour of our proposed methods.

\begin{table}[t]
    \vspace{-3ex}
    \centering
    \begin{tabular}{lllccccccc}
    \toprule
    \textbf{Position Bias Model} &~&~& $\widehat{\mathsf{P}}_{{\rm dcg}}(V|R)$ &~& $\widehat{\mathsf{P}}_{{\rm log}}(V|R)$ &~& $\widehat{\mathsf{P}}_{{\rm exp}}(V|R)$ &~& $\widehat{\mathsf{P}}_{{\rm prob}}(V|R)$ \\
         \cline{4-10}
\textbf{Relative Improvement} &~&~& 100\% &~& -3\% &~& -20\% &~& \textbf{+16\%} \\
    
\bottomrule
    \end{tabular}
    \caption{Relative correlation improvement over $\widehat{\mathsf{P}}_{{\rm dcg}}$ between DCG estimates and online metrics, higher is better.}
    \label{tab:results2}
\end{table}
\vspace{-2ex}\section{Conclusions \& Outlook}\label{sec:conclusion}
In this work, we have argued the value of a probabilistically motivated position bias model to accurately estimate exposure probabilities with a minimal number of parameters.
We have presented a specific instantiation of this general idea, modelling scroll depth via the Yule-Simon distribution, and leveraging its survival function to obtain closed-form position bias estimates.
A general approach to learn distribution parameters conditional on contextual information via Maximum Likelihood Estimation, amenable to gradient descent, allows for flexible and efficient optimisation.
Using real-world data from a social media platform, we have empirically validated that our novel methods model empirical exposure significantly more accurately than competing methods, and that the obtained estimates pave the way for improvements in unbiased offline evaluation.
In future work, we wish to extend our approach to incorporate cascade-based alternatives.\looseness=-1


\bibliographystyle{ACM-Reference-Format}
\bibliography{bibliography}

\end{document}